\documentclass[12pt]{article}

\pdfoutput=1

\usepackage{amsmath, amssymb}
\usepackage{graphicx} 
\usepackage[margin=1in]{geometry}
\usepackage{natbib}

\usepackage[utf8]{inputenc}
\usepackage[pagewise]{lineno} 

\usepackage{pdflscape} 
\usepackage{rotating} 

\usepackage{ulem} 

\usepackage{setspace} 

\usepackage[font=small,labelfont=bf, width=0.95\textwidth]{caption}
\usepackage{pdfpages} 

\usepackage{epigraph}
\title{On the use and abuse of Price equation concepts in ecology}
\author{ \normalsize{$^{a1}$Pradeep Pillai, $^{1}$Tarik C. Gouhier}\\
\\
\normalsize{$^{1}$Marine Science Center, Northeastern University,}\\
\normalsize{430 Nahant Rd, Nahant, MA 01908,}\\
\normalsize{$^{a}$To whom correspondence should be addressed. E-mail: pradeep.research@gmail.com} }

\date{}

\begin{document}

\linespread{1.0} 


{\centering

\textbf{\Large{On the use and abuse of Price equation concepts in ecology}} \\
\bigskip
Pradeep Pillai\textsuperscript{a1} , Tarik C. Gouhier\textsuperscript{1} \\
\textit{
\textsuperscript{1}Marine Science Center, Northeastern University,\\
430 Nahant Rd, Nahant, MA 01908\\
\textsuperscript{a} To whom correspondence should be addressed. E-mail: pradeep.research@gmail.com \\
}
}


%

\begin{abstract}
  \noindent In biodiversity and ecosystem functioning (BEF) research, the Loreau-Hector (LH) statistical scheme is widely-used to partition the effect of biodiversity on ecosystem properties into a ``complementarity effect'' and a``selection effect''. This selection effect was originally considered analogous to the selection term in the Price equation from evolutionary biology. However, a key paper published over thirteen years ago challenged this interpretation by devising a new tripartite partitioning scheme that purportedly quantified the role of selection in biodiversity experiments more accurately. This tripartite method, as well as its recent spatiotemporal extension, were both developed as an attempt to apply the Price equation in a BEF context. Here, we demonstrate that the derivation of this tripartite method, as well as its spatiotemporal extension, involve a set of incoherent and nonsensical mathematical arguments driven largely by na\"ive visual analogies with the original Price equation, that result in neither partitioning scheme quantifying any real property in the natural world.  Furthermore, we show that Loreau and Hector's original selection effect always represented a true analog of the original Price selection term, making the tripartite partitioning scheme a nonsensical solution to a non-existent problem. To avoid any future misuses of the Price equation in ecology, we demonstrate how its formulation can be derived by viewing it as a scheme for parsing the total change in a system into its variational and transformational components. We then show how to properly develop a BEF version of the Price equation, including a geometric interpretation that allows for easy visualization of the changes tracked by the the equation's terms. Our clarifications and new interpretive approach should help end current misunderstandings regarding `Price equation partitioning' in BEF research, and consequently, allow for a better understanding of the relationship between biodiversity and ecosystem functioning.
\end{abstract}

\noindent \textbf{Keywords}: biodiversity; ecosystem functioning; Price equation; selection effect; dominance effect; trait-dependant complementarity effect; complementarity effect; net biodiversity effect; tripartite partitioning; spatiotemporal partitioning
\bigskip

\def\undertilde#1{\mathord{\vtop{\ialign{##\crcr
$\hfil\displaystyle{#1}\hfil$\crcr\noalign{\kern1.5pt\nointerlineskip}
$\hfil\widetilde{}\hfil$\crcr\noalign{\kern1.5pt}}}}}

\section*{Introduction}
One of the central goals of ecology over the last thirty years has been to resolve the relationship between biodiversity and ecosystem functioning (BEF), specifically, how increasing biodiversity enhances the aggregate ecosystem properties (so-called ``ecosystem functioning'') observed in communities. Although the value of this research was initially challenged \citep{huston_97, kaiser_00a, huston_00}, the early controversies eventually subsided with the development of analytical tools that purportedly allowed researchers to quantify the relative contributions of various ecological factors
to the enhanced ``functioning'' often observed in diverse species mixtures and communities. The theoretical framework for analyzing BEF data ultimately culminated in the Loreau-Hector partitioning method \citep[LH partitioning;][]{loreau.hector_01}, which decomposed the total observed ecosystem change into  effects arising from two distinct categories of community-level processes: the ``selection effect'' arising from species selection or competition, and a ``complementarity effect'' associated with changes that occur from the growth of the community along the niche partitioning axis.

Loreau and Hector's selection effect quantified the increase in ecosystem properties due to species with higher than average monoculture yields dominating the community. This was  measured using the covariance between species monoculture yields, $M$, and the change in the proportion of this monoculture yield (or relative yield), $p$, that was observed in mixtures relative to an expected relative yield: $n\mathrm{Cov}[M, \Delta p]$ (where $n$ is the number of species in the mixture). The complementarity effect, on the other hand, measured ecosystem increases resulting from the average expansion of the community as a whole, and simply involved the product of the average monoculture yields and the average proportional changes, $n\overline{M}\;\overline{\Delta p }$. The Loreau-Hector partitioning was thus a simple decomposition of the total observed change in a given ecosystem property, $\phi$, into these two components
\begin{flalign}
  \Delta \phi &=  n\mathrm{Cov}[M, \Delta p] + n\overline{M}\;\overline{\Delta p }. &{}  \notag
\end{flalign}

Loreau and Hector claimed that their selection term (the covariance term above) was analogous to the selection term in the Price equation from evolutionary biology, which described the shift in the trait value of a population due to variation in trait fitness after one time step  \citep{price_70}. In a similar fashion, the Loreau-Hector selection term (LH selection) can describe how an ecosystem property will change through species competition or selection in mixtures.

Despite some early debate as to how to infer the relative contributions of selection and complementarity in the LH method \citep{petchey_03}, the LH partitioning method ultimately established itself as \textit{the} theoretical and statistical foundation for the analysis of BEF data and experiments \citep{cardinale.matulich.ea_11,cardinale.duffy.ea_12a}. However, the idea that the LH selection term meaningfully described processes analogous to natural selection was called into question in a key paper published over a decade ago in \textit{Ecology Letters} \citep{fox_05}. This paper argued that the effect quantified by the LH selection expression is not analogous to the selection effect measured by the Price selection term because the LH selection term does not describe changes that operate as a``zero-sum game''. In response to these perceived limitations, Fox \citeyearpar{fox_05} created a new arithmetic expression called the ``dominance effect'' to more accurately capture and measure the effect of selection in BEF experiments. This new dominance effect was then subtracted from the original LH selection effect to create yet another new quantity, the ``trait-dependent complementarity effect''.

These two new effects -- the dominance and trait-dependent complementarity effects -- combined with the original complementarity effect, which \citet{fox_05} renamed the ``trait-independent complementarity effect'', together constitute the tripartite method for partitioning the effects of biodiversity on ecosystem functioning. Fox \citeyearpar{fox_05}'s original critique and his corrections to the perceived limitations of the LH selection effect are now accepted as a standard component of BEF analysis in ecology. Indeed, the tripartite method has been used extensively to analyze BEF experimental data for nearly a decade-and-a-half \citep{bruno.lee.ea_06, long.bruno.ea_07, fox.rauch_09, pires.srivastava.ea_18}. Additionally, it has served as the basis for statistical and theoretical models \citep{baert.jaspers.ea_17,baert.eisenhauer.ea_18}, and it has been extended in a spatiotemporal manner \citep{isbell.cowles.ea_18}.

We demonstrate below that, in spite of its widespread use, this entire statistical methodology was ill-conceived, and much of the theoretical and experimental works based on it, unfortunately misdirected. The two key measures derived from the tripartite method -- the dominance and the trait-dependent complementarity effects -- are nonsensical measures in that they do not actually quantify any meaningful or discernible property in the natural world. This holds even more so for the new measures created by the spatiotemporal extension of the tripartite approach \citep{isbell.cowles.ea_18}. It is important to note that the problematic nature of the tripartite partitioning, as well as its spatiotemporal extension \citep{isbell.cowles.ea_18}, are not due to a simple set of methodological or mathematical errors. Rather, they arise from a whole host of incoherent claims and arguments based on na\"ive analogies with the Price equation from evolutionary theory that have somehow gone unnoticed in ecology and biodiversity studies for more than a decade.


Much of the reasoning that led to the misuse of the Price equation in BEF research can be traced back to the \textit{prima facia} absurd (and now standard) claim that the Loreau-Hector selection effect does not operate as a ``zero-sum game'', and thus cannot constitute a true selection effect, since  ``[in] any process analogous to evolution by natural selection [...] differences in the frequency of `parents' and their `offspring' will sum to zero, so that  $\mathbb{E}[\Delta q]=0$ [for a given trait frequency $q$]'' \citep{fox_05}.

Of course this is clearly false. The property $\mathbb{E}[\Delta q]=0$ will \textit{always} hold when comparing \textit{any} two sets of objects (whether related or not) at any time and place, given that it is simply a trivial arithmetic consequence of relative frequencies (herein referred to simply as frequencies) always summing-up to one, or having an average of $1/n$ (for $n$ total elements). Hence, this is not a particular characteristic of processes ``analogous to evolution by natural selection''. As such, any change, including that described by the Loreau-Hector selection effect, will exemplify this property; what determines whether a change represents a zero-sum game, on the other hand, will be determined by the particular set of constraints it is under.

Below we demonstrate that in the Price equation it is the fitness or expected growth rates of individual species that defines the constraint under which the zero-sum game of selection is operating. Additionally, we show how Loreau and Hector's selection effect can be viewed as a similar type of constrained change. Furthermore, using a vector-based geometric approach we demonstrate how this constraint in BEF studies can be visualized as a space in the form of a simplex connecting all monoculture yields, within which ecosystem shifts will represent changes due to selection such that species gains come at the expense of others in a zero-sum manner.

After these initial claims regarding ``zero-sum games'', the subsequent mathematical arguments used to develop the tripartite method only become curiouser and curiouser. Much of the argumentation for the tripartite method and its spatiotemporal extension were largely justified by a misguided attempt to apply the Price equation in a BEF context. Surprisingly for such a simple identity, the Price equation has been the source of so much confusion in ecology. To help demystify the Price equation and avoid such abuses in the future, we will explore both the equation, in general, and the selection effect, in particular.

We we will first demonstrate how the classic Price equation from evolutionary biology can be effortlessly derived by viewing it as a simple identity that decomposes the total change observed in a system into the sum of what can be considered its \textit{variational} and \textit{transformational} components  \citep[][]{lewontin_85}. Based on our derivation, we will then demonstrate how a serious misunderstanding of what the Price equation is actually measuring has led to both to the misguided criticism of the LH selection term and the subsequent development of the tripartite method. Finally, we will show how one can develop a true BEF version of the Price equation using a vector-based geometric approach. This geometric approach will help provide a visually intuitive understanding of the different changes quantified by the terms of the BEF Price equation, and in doing so, illustrate how the original LH selection term has always described a type of selection effect, at least under the assumption of linear ecosystem function-abundance relationships \citep{pillai.gouhier_18}.

Despite its previous misuses, the Price equation may still yet serve a positive --- albeit more modest --- role in ecosystem research by allowing investigators to be more explicitly aware of the assumptions and expectations inherent in their experimental designs. In this manner the Price equation may offer a useful conceptual tool for exploring the forces driving ecosystem changes.

\section*{The Price equation}
\subsection*{Variational and transformational change}
The Price equation is ultimately a way of tracking how aggregate properties in a system change over time based on the expected change or growth of the system's various components that bear those properties. The system components can be any set of entities or groups (e.g., species, lineages, genes) that can grow independently of each other over time, while the properties of interest (e.g., biomass, carbon fixation, average trait value) are the attributes that are associated with those entities. In BEF studies the groups are often species and the properties refer to so-called ``ecosystem functions'' like biomass.

Let us say we have $n$ entities or groups, and that $\phi$ represents the initial value of the total property of interest, while $\phi_i$ represents the property value contributed to this total by the \textit{i}th entity. Our system's total or aggregate property value would then be $\phi =\sum^n_i \phi_i$. Now if we assume that each entity $i$ on its own is expected to grow by a factor of $w_i$ in a time step, then we could predict the expected change in the aggregate system property due to collective growth of all groups as  $\phi_{_\mathbf{V}} =\sum^n_i w_i \phi_i$.  We will refer to this multiplicative factor, $w_i$, as the fitness. The actually observed aggregate property, $\phi_{_\mathbf{obs}} =\sum^n_i \phi'_i$ (where $\phi'_i$ is the observed contribution of entity $i$) might, however, be different from that expected due to variation in growth, $\phi_{_\mathbf{obs}} \neq \phi_{_\mathbf{V}}$.

The total change in the aggregate property value of the system just described can be considered as arising due to effects associated with \textit{variational} and \textit{transformational} evolution or shifts in the system \citep[\textit{sensu amplo,}][]{lewontin_85}. The property effect expected due to variation in individual growth rates (\textit{variational} evolution) is simply $\phi_{_\mathbf{V}}=\sum^n_i w_i \phi_i$. The difference in the system property between this expected value and that of the observed state $\phi_{_\mathbf{obs}}=\sum^n_i \phi'_i$, represents an additional shift due to the transformational changes in the system's aggregate property that are not simply reducible to variation in growth rates (in BEF studies this could be due to ecological interactions). This gives the transformational change as $\Delta\phi_{_\mathbf{T}} = \phi_{_\mathbf{obs}} - \phi_{_\mathbf{V}}$, which can be rewritten as $\Delta\phi_{_\mathbf{T}} = \sum^n_i w_i (\frac{\phi'_i}{w_i} - \phi_i)$.

The expression below is a modified form of the Price equation that shows how the total observed property can be partitioned into the effects of variational and transformational change:
\begin{flalign} \label{eq:Price1}
\phi_{\mathbf{obs}} &= \phi_{_\mathbf{V}} + \Delta\phi_{_\mathbf{T}}. &{}
\end{flalign}
This can be rewritten using the expected value operator as
\begin{flalign} \label{eq:Price2}
\phi_{\mathbf{obs}} &= n\mathbb{E}[w \phi] + n\mathbb{E}[w \Delta\phi], &{}
\end{flalign}
where $\Delta \phi_i = (\widehat{\phi'_i} - \phi_i) = (\frac{\phi'_i}{w_i} - \phi_i)$ is the change in the scaled or average observed property value of the \textit{i}th group or species.

The classic Price equation itself is usually only concerned with a portion of the total variational change in the system. Using the mathematical expression for the expected value of the product of two random variables $X$ and $Y$, $ \mathbb{E}[XY] = \mathrm{Cov}[X,Y] + \mathbb{E}[X]\mathbb{E}[Y]$, we can partition the total variational change in Eq. \eqref{eq:Price2} as follows: $ n\,\mathbb{E}[w \phi] = n\mathrm{Cov}[w,\phi] + n\,\overline{w}\,\overline{\phi}$. The Price equation in its standard form only considers the portion of the total variational change that is described by the covariance term, what we will refer to here as the ``variational selection'' term which describes the shift in the relative property contributions of each entity or species due to the effect of selection.

The Price equation expressed as the \textit{total} change in property can be obtained by subtracting the average expected growth of the system, $n\,\overline{w}\,\overline{\phi}$, from both sides of Eq. \eqref{eq:Price2},
\begin{flalign} \label{eq:Price3}
 \sum_{i=1}^n \left( {\phi'_i}  -  \overline{w} \, \phi_i \,\right)  &= n\mathrm{Cov}[w,\phi]+ n\mathbb{E}[w \Delta\phi]. &{}
\end{flalign}
Dividing by the number of groups, $n$, gives the version of the Price equation describing the per group change in the property value, which, with some rearrangement of the LHS, can be written as
\begin{flalign} \label{eq:Price4}
 \overline{w}\, \sum_{i=1}^n \left( \frac{w_i}{\sum_j w_j} \; \widehat{\phi'_i}  -  \frac{1}{n} \, \phi_i \,\right)  &= \mathrm{Cov}[w,\phi]+ \mathbb{E}[w \Delta\phi], &{}
\end{flalign}
where ${w_i }/{\sum_j w_j}$ (which can be represented as $f'_i$) gives the final observed frequency weight of the $i$th group's property value, and $\widehat{\phi'_i} = {\phi'_i }/{w_i}$ is the observed property contribution of the $i$th group scaled by its growth rate. This scaled value, $\widehat{\phi'_i}$, can also be interpreted as the \textit{average} observed property value of group $i$, particularly when $i$ is comprised of smaller discrete units capable of growing in number (such as a species $i$ composed of self-reproducing individuals). The LHS of the above equation, $\overline{w}\, \sum_i (f'_i \, \widehat{\phi'_i} -\frac{1}{n}\,\phi_i)$, shows how the per group change in the property can be expressed as the product of the average fitness $\overline{w}$ and the change in the \textit{scaled} or average property in the system, $\Delta \overline{\phi}$, allowing us to express the Price equation in its standard form:
\begin{flalign} \label{eq:Price}
\overline{w}\, \Delta \overline{\phi}   &=  \mathrm{Cov}[w,\phi]+ \mathbb{E}[w \Delta\phi]. &{}
\end{flalign}
Expressions \eqref{eq:Price1}-\eqref{eq:Price} demonstrate how the Price equation in its general form ultimately represents the rearrangement of a simple mathematical identity for the expected value of a product of two variables.

Since the transformational term, $\Delta \phi_{_\mathbf{T}}$, represents observed departures of the aggregate property value from that expected based on the growth rate of groups or species, it is used in BEF research to measure the \textit{net biodiversity effect}, which purportedly quantifies the effect community diversity has on ecosystem properties. Hence, it is the transformational term that is of interest in BEF studies, and it is this term that the Loreau-Hector method partitions in an attempt to infer underlying causal forces driving diversity-ecosystem functioning relationships (Loreau and Hector 2001, Pillai and Gouhier, 2018).

As we will demonstrate in detail later, the LH method in BEF studies can be viewed as a further partitioning of the transformational term, $n\,\mathbb{E}[w \Delta\phi]$, where the ecosystem property of interest is represented as a proportion of the monoculture yield (i.e., the relative yield), $p_i$, and where monoculture yields serve as a proxy for expected growth or fitness, $w_i=M_i$:
\begin{flalign} \label{eq:LH}
n\,\mathbb{E}[M \Delta p] &= n\mathrm{Cov}[M,\Delta p] + n\,\overline{M}\,\overline{\Delta p}. &{}
\end{flalign}

Loreau and Hector \citeyearpar{loreau.hector_01} claim that the covariance term, $n\mathrm{Cov}[M,\Delta p]$, obtained above by partitioning the transformational effect also constitutes a ``selection effect'' similar to the variational selection effect. Therefore, for convenience, we will refer to the LH selection term, $ n\mathrm{Cov}[M,\Delta p]$, as the ``transformational selection'' term. Fox's \citeyearpar{fox_05} claim that LH selection does not constitute a true selection effect because it does not describe a zero-sum game is, in part, born of a confusion between what can be considered the variational selection and the transformational selection terms of the Price equation in BEF studies.

\subsection*{Selection effect and the arithmetic properties of the covariance operator}
Since there is so much confusion over the selection effect in the Price equation, we now explore the variational selection term in more detail. Let us assume that a system has $n$ species, where each of the $i$ species in the community is composed of $x_i$ individuals. Then, in total, the community will have $n$ species and $N$ individuals, where $N=\sum_{i=1}^n x_i$. Under the assumption that the property of a species is a linear function of its abundance (an assumption we will make throughout, unless otherwise stated), the total property value of the $i$th species, $\phi_i$ will be
\begin{flalign}
  \phi_i &= a_i x_i, &{}
\end{flalign}
where $a_i$ is the per capita property value for species $i$.

Since we will focus, for the time being, only on the variational component of the Price equation, we can claim that the change observed in property $\phi_i$ after a single time step for all species will be due to the variation in growth rate that each species would otherwise be expected to show on its own (i.e., independent of each other), $w_i$. The total property of the system after a time step due to variational growth of all species, $ \phi_{_\mathbf{V}}$, is then
\begin{flalign}\label{eq:price_var}
  \phi_{_\mathbf{V}} &= \sum_{i=1}^n w_i \phi_i = \sum_{i=1}^n w_i (a_i x_i). &{}
\end{flalign}
The total community property $\phi_{_\mathbf{V}}$ after a single time step is found in Eq. \eqref{eq:price_var} by summing over $n$ species. However, Eq. \eqref{eq:price_var} can be rewritten so that $\phi_{_\mathbf{V}}$ is summed over $N$ individuals,
\begin{flalign}
  \phi_{_\mathbf{V}} &= \sum_{i=1}^n w_i (a_i x_i) =  \sum_{j=1}^N w_j a_j. &{}
\end{flalign}
This seemingly innocuous shift in how the terms are summed will have a critical impact on the way total system change will be partitioned, as we will see shortly. Now since $\sum_{j=1}^N w_j a_j = N\times\mathbb{E}[wa]$, and given the mathematical identity $ \mathbb{E}[YZ] = \mathrm{cov}[Y,Z] + \mathbb{E}[Y]\mathbb{E}[Z]$ for any two random variables $Y$ and $Z$,  we can rewrite $\phi_{_\mathbf{V}}$ as
\begin{flalign}\label{eq:variational}
  \sum_{j=1}^N w_j a_j &= N\,\mathrm{Cov}[w, a] + \overline{w}\sum_{j=1}^N a_j &{}
\end{flalign}
If $x_i'$ represents the abundance of species $i$ after a time step, then the expected growth rate (or fitness) of species $i$ is $w_i=x_i'/x_i$. We can then express the expected value of the growth or fitness, $\mathbb{E}[w]= \overline{w}$, as
\begin{flalign} \label{eq:fitness}
   \overline{w} &=  \frac{1}{N} \sum_{j=1}^N w_j =  \frac{1}{N} \sum_{i=1}^n w_ix_i = \frac{\sum_{i=1}^n w_i x_i}{\sum_{i=1}^n x_i}. &{}
 \end{flalign}

As mentioned, the normal form of the Price equation is usually expressed by considering only the covariance component of the total variational change, $\phi_{_\mathbf{V}} = \sum_j^Nw_ja_j =N\mathrm{Cov}[w, a]+ N\overline{w}\overline{a}$. The portion of the variational change that is encompassed by the covariance term is referred to as the ``selection'' term of the Price equation, which can be obtained by rearranging Eq. \eqref{eq:variational} as follows,
\begin{flalign}\label{eq:selection_total}
  N\mathrm{Cov}[w, a] &=  \sum_{j=1}^N w_j a_j - \overline{w}\sum_{j=1}^N a_j, &{} \notag \\
  &= \overline{w} \times  \sum_{j=1}^N \left( \frac{w_j}{\overline{w}} -1 \right) a_j. &{}
\end{flalign}
Eq. \eqref{eq:selection_total} represents the \textit{total} shift due to selection of the system property away from that expected from the average growth rate. The \textit{per capita} change in the system due to selection can be found by dividing \eqref{eq:selection_total} by $N$,
\begin{flalign}\label{eq:selection_per_cap}
  \mathrm{Cov}[w, a] &=   \overline{w} \times  \sum_{j=1}^N \left( \frac{w_j}{\sum_k^N w_k} - \frac{1}{N} \right) a_j. &{}
\end{flalign}
Or alternatively, the selection term can be expressed as the change in the \textit{average} or scaled property value by dividing \eqref{eq:selection_per_cap} by $\overline{w}$ (so long as $\overline{w}\neq0$):
\begin{flalign}\label{eq:selection_ave}
  \frac{1}{\overline{w}} \times \mathrm{Cov}[w, a] &=     \sum_{j=1}^N \left( \frac{w_j}{\sum_k^N w_k} - \frac{1}{N} \right) a_j. &{}
\end{flalign}
The above three expressions represent the Price equation when there is no transformational component (i.e., $\mathbb{E}[\Delta \phi \; w]=0$).

Recalling from \eqref{eq:fitness} that $\overline{w}= \frac{1}{N} \sum_{i=1}^n w_ix_i$, we can rewrite the expression for Eq. \eqref{eq:selection_ave} by summing the RHS over the $n$ species instead of $N$ individuals,
\begin{flalign}
  \frac{1}{\overline{w}} \times \mathrm{Cov}[w, a] &=     \sum_{i=1}^n \left( \frac{w_i x_i}{\sum_k^n w_k x_k} - \frac{x_i}{\sum_k^n x_k} \right) a_i, &{} \notag
\end{flalign}
which, when we substitute $f_i$ and $f'_i$ for species $i$'s' fraction of the total abundance before and after the time step, becomes the following
\begin{flalign}\label{eq:selection_fraction}
  \frac{1}{\overline{w}} \times \mathrm{Cov}[w, a] &=     \sum_{i=1}^n \left(f'_i - f_i \right) a_i, &{}
\end{flalign}
The RHS of Eq. \eqref{eq:selection_fraction} can be written as $\Delta \overline{a} = \overline{a'} - \overline{a}$; that is, the difference in the average \textit{per capita} property of the system before and after a single times step. This allows us to express the selection term of the Price equation (without the transformational term) as $\overline{w} \Delta \overline{a} = \mathrm{Cov}[w, a]$ (or alternatively, $\Delta \overline{a} = \frac{1}{\overline{w}}\mathrm{Cov}[w, a]$). If, as in population genetics, we replace per capita property $a_i$ by trait value $z_i$, we get the more familiar expression $\overline{w} \Delta \overline{z} = \mathrm{cov}[w, z]$.

Considering also that $\sum_{i=1}^n  \Delta f_i \; a_i, = n\,\mathrm{Cov}[\Delta f, a]$, we then have multiple ways of representing the Price equation's average selection effect on the \textit{per capita} property value $a_i$:
\begin{flalign}\label{eq:selection_price}
  \frac{1}{\overline{w}} \times \mathrm{Cov}[w, a] &= \sum_{i=1}^n  \Delta f_i \; a_i, = n\,\mathrm{Cov}[\Delta f, a]. &{}
\end{flalign}

The above expression \eqref{eq:selection_price} represents the selection effect when the covariance is computed between fitness and the \textit{per capita} or \textit{individual} property value. We could also have measured another selection effect relative to the \textit{per species} property value by partitioning the variational change $\phi_{_\mathbf{V}}$ as the sum across $n$ species, $\sum_{i=1}^n w_i \phi_i$, instead of across $N$ individuals, $\sum_{i=1}^N w_i a_i$ as seen in Eq. \eqref{eq:price_var}. Both approaches (or any other representation and partitioning of $\phi_{_\mathbf{V}}$) would have been just as legitimate depending on the biological question of interest. However, although both total sums in \eqref{eq:price_var} are equal quantities, their respective covariance terms are not. That is, the total change represented is the same, but how this total is decomposed or redistributed into components may not be ($\sum_{i=1}^n w_i \phi_i = \sum_{i=1}^N w_i a_i$ but $\mathrm{Cov}[w, a]\neq \mathrm{Cov}[w, \phi]$).

\section*{The tripartite method}
\subsection*{Comparing the Price and the Loreau-Hector selection effects}
Expression \eqref{eq:selection_price} is at the crux of the argument that Loreau and Hector's selection effect is not a true ``selection'' effect. In \eqref{eq:selection_price} the variational selection effect expressed as the \textit{scaled per capita} property value can be written as $\frac{1}{\overline{w}} \, \mathrm{Cov}[w, a] = n\,\mathrm{Cov}[\Delta f, a]$.  In the LH partitioning (Eq. \eqref{eq:LH}), the transformational selection term describing the \textit{unscaled per species} changes in property value is given by $n\mathrm{Cov}[M,\Delta p]$.

In Fox \citeyearpar{fox_05} the unscaled Loreau-Hector (transformational) selection term, $n\mathrm{Cov}[M, \Delta p]$, was directly compared to the scaled or average version of the Price selection term when rewritten or expressed as  $n\mathrm{Cov}[\Delta f, a]$, imagining that $\Delta p$ was comparable to $\Delta f$. By making what appears to be an argument based purely on a \textit{visual analogy} between the two expressions, \citet{fox_05} originally claimed that the LH selection term does not fit the criteria of a proper selection term because $p$ is not a measure of the frequency or proportion of the total of all groups or species, like $f$ is in \eqref{eq:selection_price}. This is an astonishing statement based not on mathematical reasoning but on a facile visual comparison of two mathematical expressions.

Indeed, no attempts appear to have been made to consider what the two expressions, $\Delta p$ and $\Delta f$, describe biologically, and then \textit{mathematically} find any of the points of congruence or dissimilarity between them. A couple of issues immediately present themselves: the first is that there seems to be confusion about what plays the corresponding role of `fitness' in the Loreau and Hector selection effect, or more importantly, it does not appear to matter. We will address this in a moment. The other issue is that $f$ and $p$ need not be comparable: $f_i$ in Eq. \eqref{eq:selection_price} represents the \textit{frequency} of a component or type $i$ in the system, whereas $p_i$ is simply the \textit{state} of component $i$ (the state here being represented as the proportion of species $i$'s total monoculture yield). As such, the $p_i$ values across all species are \textit{not} required to sum to unity, as is the the case with the frequencies $f$.

Furthermore, a frequency $f$ will always appear when one rearranges the Price selection term so as to express it as an average per group change, as opposed to the total change due to selection represented by Eq. \eqref{eq:selection_total} (or Eq. \eqref{eq:selection_per_cap}). What is important to note is that the \textit{functional form} of the LH selection term, $n\times\mathrm{Cov}[\Delta p, M]$, is actually comparable, not to the \textit{scaled} expression in \eqref{eq:selection_ave}, but to the total selection effect in Eq. \eqref{eq:selection_total}, $N\times\mathrm{Cov}[w, a]$.

In the exact same way that the total change due to selection in the classic Price equation can be rearranged to express the change  in frequency or fraction of a property relative to the average, $\frac{1}{\overline{w}}\mathrm{Cov}[w, a] = n\mathrm{Cov}[\Delta f, a]$, the LH selection term can similarly be rearranged, first by starting with the definition of the covariance,
\begin{flalign}
n\mathrm{Cov}[\Delta p, M] &= \sum_{i=1}^n  \Delta p_i\; M_i -  n\overline{\Delta p}\; \overline{M},   &{}
\end{flalign}
and then by rearranging terms,
\begin{flalign}\label{eq:lh_cov}
\frac{1}{\overline{\Delta p}} \mathrm{Cov}[\Delta p, M] &= \sum_{i=1}^n  \left(\frac{\Delta p_i}{\sum_{j=1}^n \Delta p_j} - \frac{1}{n} \right) M_i &{} \notag \\
&= \sum_{i=1}^n  \left(f_i'-\frac{1}{n} \right) M_i &{}  \notag \\
&= n\mathrm{Cov}[\Delta f,M]. &{}
\end{flalign}
Here the final frequency or fraction $f'_i$ now represents $f'_i= {\Delta p_i}/{\sum^n_j \Delta p_j}$. As we can see, the LH selection term can be expressed in terms of fractions of a total that sums to unity, just as was done in Eq. \eqref{eq:selection_ave}, and via the same method as well. Clearly this is not a point of distinction between the Price selection and LH selection terms.

Even more striking is the false claim that the Price selection term, when rewritten in the form of Eq. \eqref{eq:selection_ave}, represents the operation of selection or some analogous process, and of what is referred to as a ``zero-sum game''. The fact that the Price selection term can be written in the form of Eq. \eqref{eq:selection_ave} is due to a general and simple arithmetic property of the covariance operator, and is not a specific feature of cases involving evolution or selection-like processes. Indeed the covariance between \textit{any} two random variables $Y$ and $Z$ can be written in the same fashion:
\begin{flalign} \label{eq;covariance_eg}
  \frac{1}{\overline{Y}} \mathrm{Cov}[ Y, Z] = \sum_{i=1}^n  \left(\frac{Y_i}{\sum_{j=1}^n Y_j} - \frac{1}{n} \right)Z_i  &= \sum_{i=1}^n  \left(f_i'-\frac{1}{n} \right) Z_i &{} \notag \\
   &= n\mathrm{Cov}[\Delta f,Z]. &{}
\end{flalign}

In short, claiming that $n\mathrm{Cov}[\Delta p, M]$ is not a true ``selection effect''  because the sum of all $p_i'$ (or $p_i$) does not sum to one is as absurd as claiming that $\mathrm{Cov}[w, a]$ in the classic Price equation cannot represent a change due to selection because neither of the variables $w_i$ or $a_i$ are frequencies that add up to one.

\subsection*{The ``Dominance effect''}
In order to ``correct'' the perceived limitations of the LH selection term Fox \citeyearpar{fox_05} developed the ``dominance effect''. This was done by literally forcing the LH selection term to \textit{visually} resemble Eq. \eqref{eq:selection_price}, specifically $\frac{1}{\overline{w}} \mathrm{Cov}[w, a] = \sum_{i=1}^n  \Delta f_i \; a_i$. This was achieved by dividing the final and initial proportions of monocultures, $p'_i$ and $p_i$, respectively, by the total proportions $\sum_k p'_k$ and $\sum_k p_k$. Basically, to use the Price selection term written in the form seen in \eqref{eq:selection_fraction},
\begin{flalign} \label{eq:dominance_template}
\sum_{i=1}^n (f'_i -f_i)a_i &=     \sum_{i=1}^n \left( \frac{w_i x_i}{\sum_k^n w_k x_k} - \frac{x_i}{\sum_k^n x_k} \right) a_i, &{}
\end{flalign}
as a visual template to modify the LH selection term and create
\begin{flalign} \label{eq:dominance_general}
\sum_{i=1}^n (f'_i -f_i)M_i &=     \sum_{i=1}^n \left( \frac{p'_i }{\sum_k^n p'_k} - \frac{p_i}{\sum_k^n p_k} \right) M_i. &{}
\end{flalign}

Eq. \eqref{eq:dominance_general} is an attempt to force the LH selection term to appear in the form of Eq. \eqref{eq:dominance_template}. One immediate problem with this forcing based on a superficial visual analogy is that the argument that expression \eqref{eq:dominance_general} is analogous to \eqref{eq:dominance_template} will only hold if ${p_i}/\,{\sum_k^n p_k} = {1}/{n}$ holds for all $p_i$. Since the standard experimental design in BEF studies assumes the initial proportion $p_i$ is $1/n$ for all species, the above condition holds in practice, and as a result this problematic limitation (indicative of yet deeper issues) is not dealt with (or even considered as problematic) any further. In other words, \eqref{eq:dominance_general} is not, in fact, comparable in a general sense to the Price selection. The reason expression \eqref{eq:dominance_template} appears to allow the initial proportion of individuals to differ from $1/n$ is because \eqref{eq:dominance_template} involves switching the index of summation from $j$ individuals to $i$ species, such that $1/N$ can be rewritten as $x_i/N$; an analogous index-switching is not possible with \eqref{eq:dominance_general}, which involves the proportion of monoculture, $p$, as opposed to count data.

This improvised transformation of LH's selection term into Eq. \eqref{eq:dominance_general} when $p_i =1/n$ represents the so-called ``dominance effect'' \citep{fox_05}:
\begin{flalign} \label{eq:dominance}
\textrm{Dominance Effect} &\overset{\underset{\mathrm{def}}{}}{=} \sum_{i=1}^n \left( \frac{p'_i }{\sum_k^n p'_k} - \frac{p_i}{\sum_k^n p_k} \right) M_i &{} \notag \\
&= \sum_{i=1}^n \left( \frac{p'_i }{\sum_k^n p'_k} - \frac{1}{n} \right) M_i \;\;\;\;\text{   (when $p_i=p$,\;\;$\forall i$)}. &{}
\end{flalign}
It is important to note that the $\frac{1}{n}$ in the above definition of the dominance effect, which is the initial state of each species given as a proportion of monoculture yields, is \textit{not} the same as the term $\frac{1}{n}$ (or $\frac{1}{N}$) that appears when one scales the Price equation selection expression by fitness, as seen in Eq. \eqref{eq:selection_ave}.  When one scales the selection effect (or, for that matter, any covariance operator by one of its variables, as in Eq. \eqref{eq;covariance_eg}), the term $\frac{1}{n}$ that appears will represent the \textit{average frequency} or weight for the per group property or trait value, and not the initial proportion of the property itself. This critical flaw has gone unnoticed because BEF experimental designs typically assume that the initial proportions of all species are equal to the average, which is $\frac{1}{n}$. This means that the dominance effect as defined above is not even a superficial recreation of the selection effect.

More problematic still is that the expression for the dominance effect was simply pulled out of thin air. At no point was there an attempt to derive this expression from first principles, or from the total change occurring in the system, or to work backward from the expression to see how it relates mathematically to the LH selection effect.  Instead the dominance effect was created  \textit{ex nihilo} to simply ``look like'' the scaled selection effect in evolutionary biology.

\subsection*{What exactly is the dominance effect?}
To understand what this dominance effect really represents, one can rearrange Eq. \eqref{eq:dominance} to express it as
\begin{flalign} \label{eq:dominance_cov}
\sum_{i=1}^n \left( \frac{p'_i }{\sum_k^n p'_k} - \frac{1}{n} \right) M_i &= \frac{1}{\overline{p'}} \; \mathrm{Cov}[p', M]. &{}
\end{flalign}
This shows that the dominance effect is simply the covariance between the final observed monoculture proportion and the monoculture yield, scaled or divided by the average of the observed proportion, $\overline{p'}$.

Furthermore, since the dominance effect only holds when initial monoculture proportions are all equal, it can be shown that $\mathrm{Cov}[p', M] = \mathrm{Cov}[\Delta p, M]$. We can thus rewrite the dominance effect as:
\begin{flalign} \label{eq:dominance_LH}
\textrm{Dominance Effect} &= \frac{1}{\overline{p'}} \; \mathrm{Cov}[M, \Delta p]. &{}
\end{flalign}
This definition means that the total (unscaled) ecosystem change due to the so-called ``dominance effect'',  $n\mathrm{Cov}[p', M]$, is \textit{identical} to the Loreau-Hector selection effect.

Recall that the change described by the LH selection term, $n\,\mathrm{Cov}[M, \Delta p]$, was comparable to the \textit{total} change due to variational selection in the standard Price equation as seen in Eq. \eqref{eq:selection_total}, $N\,\mathrm{Cov}[w, a]$. And like the Price selection effect, an average per capita version of the LH selection can be expressed by scaling or dividing $\mathrm{Cov}[M, \Delta p]$ by the average growth or fitness. However, from Eq. \eqref{eq:dominance_LH} we can see that although the dominance effect is a scaled version of the LH selection effect, it is \textit{not} analogous to the average variational selection term in the Price equation we saw in Eq. \eqref{eq:selection_ave}, because dividing by $\overline{p'}$ is not equivalent to dividing by the average fitness, or the average change in the size of the community. In other words, the dominance effect was never anything more than a version of the Loreau-Hector selection effect that had been arbitrarily divided, without any clear biological motivation, by $\sum_i{p_i'}$.

\subsection*{A note on the role of fitness in the dominance effect}
Earlier, we put aside the fact that the role of fitness or the expected growth in the LH selection term appears to have been largely ignored in the development of the tripartite method. This is odd since, if nothing else, the Price equation is a formula for partitioning change in a system based on (or relative to) variation in the \textit{expected} growth rates of the system's different components. This seems to be utterly lost on so much of the ecological literature that treats the phrase ``Price equation partitioning'' as synonymous with using a covariance operator.

As we just showed, the dominance effect is nothing more than a form of the LH selection effect divided by $\overline{p'}$. Dividing a covariance term by the average of either of its variables will result in two different quantities. In the Price equation, dividing the covariance term by the average fitness is useful because it allows us to control for the the average growth (expansion or contraction) of the system as a whole, so that we are left with a measure of how the average property in the system is shifting in one time step. This means that in order to reasonably measure change due to fitness, one should only scale the covariance by average fitness or the expected growth rate, and not by the average of just any variable or quantity.

As we mentioned earlier (and as we will explain in the next section), the LH selection term is best understood as arising from the partitioning of the transformational component of the Price equation, with monoculture yields giving the \textit{expected} growth factor of each species independent of all others, i.e., the fitness. Treating $p'$, which is the final observed proportion, or even the observed growth factor $w=p'/p$, as though it corresponded to fitness in the Price equation is problematic since both deal with the changes that have been observed in the mixture, and not the changes each species is expected to undergo independent of each other. However, in order to demonstrate how it fails even on its own terms, we will follow the tripartite method's implicit line of reasoning by interpreting the LH expression as though  $p'$ corresponded to the expected growth or fitness.

Now, if for argument's sake, we assume that the LH transformational selection term $n\mathrm{Cov}[\Delta p, M]$  should be directly compared to the variational selection term, and that the role of fitness can be played by the growth actually observed in mixtures $w_i= p'_i/p_i$ (i.e., growth due to transformational changes), we can derive a proper BEF transformational analog to the scaled variational selection effect in evolutionary biology. The total observed change in mixture can be represented as
\begin{flalign} \notag
\phi_{\mathbf{obs}} &= \sum_{i=1}^n p'_i M_i = \sum_{i=1}^n w_i \left( p_i\,M_i\right). &{}
\end{flalign}
The total change can be expressed in terms of the fitness of each species defined as $w_i= p'_i/p_i$. We can, after some rearrangement, obtain the portion of this total change that is due to selection, $n\mathrm{Cov}[w,(pM)]$. When (and only when) the initial proportions of all species, $p_i$, are equal ($p_i=p$), then this total selectional change can be scaled by dividing by fitness to give,
\begin{flalign} \label{eq:true_analog}
    \frac{1}{\overline{w}} \; \mathrm{Cov}[w,(pM)] &= \sum_{i=1}^n \left( \frac{p'_i }{\sum_k^n p'_k} - \frac{1}{n} \right) (pM_i).  &{}
\end{flalign}

If we keep to the implicit and problematic assumptions associated with the tripartite method \citep{fox_05}, then it is Eq. \eqref{eq:true_analog} that represents the closest BEF analog of the scaled variational selection effect from evolutionary biology, and not the visually improvised and \textit{ad hoc} dominance effect. The dominance effect, as defined in Eq. \eqref{eq:dominance_cov}, will only equal the above Eq. \eqref{eq:true_analog} when $p=1$ holds as the initial proportion for all species.

In short, even if we ignore the question regarding what parameters are supposed play the role of fitness in the dominance effect, the effect itself turns out to have never been a measure of selectional change comparable to the Price selection effect in evolutionary biology -- the very justification for its creation in the first place.

\subsection*{The trait-dependent complementarity effect}
The final strange component in the tripartite method is the step to create the ``trait-dependent complementarity'' effect. Not only was the dominance effect created without the use of any recognizable form of mathematical derivation or argument, but -- even more puzzlingly -- this dominance effect term was then subtracted from the total change in yield, $\Delta \phi_{_\mathbf{T}}$, to produce both the original Loreau-Hector complementarity effect, and a new expression dubbed the ``trait-dependent complementarity effect''. This whole step to create the trait-dependent complementarity effect ultimately rested on computing the difference between two incommensurable properties, specifically, $\phi'_i-{\phi'_i}/{\sum_jp'_j}$ -- the latter term in the expression representing a completely new and enigmatic quantity \citep[see][Eq. 10]{fox_05}.

The overall approach was to insert terms into the expression for the total yield change that would produce an expression that visually resembled what a selection term \textit{should} look like. The end result was that the trait-dependent complementarity effect could be defined as the remainder left over after the dominance effect was subtracted from the LH selection effect.

What is confusing in this final step to create the trait-dependent complementarity effect is the arbitrary subtraction of a relative or \textit{scaled} quantity of change, like the dominance effect, from  an \textit{absolute} quantity of change like the LH selection effect, or alternatively, from the total yield, $\Delta \phi$. Because the dominance effect is not immediately comparable to a quantity like the LH selection term, the creation of the trait-dependent complementarity effect essentially involves the subtraction of two non-comparable quantities:
\begin{flalign} \label{eq:trait_ind_complement}
\textrm{Trait-dependent complementarity} &= n\mathrm{Cov}[\Delta p, M] -   \left(\frac{1}{\overline{p'} } \right)  \, \mathrm{Cov}[\Delta p, M]. &{}
\end{flalign}

At no point in the \textit{ad hoc} argument for the trait-dependent complementarity effect could we find a mathematical or biological justification for this final step, only a \textit{post hoc} verbal rationalization. The absurdity of creating a new quantity based on the difference between a relative quantity and its corresponding absolute quantity is further underscored when one asks why the same approach is not similarly applied to the selection term in the original trait-based evolutionary Price equation, $n\mathrm{Cov}[w, z]$; specifically by creating a new ``trait-dependent'' quantity defined by $ n\mathrm{Cov}[w, z]  - \frac{1}{\overline{w}}\mathrm{Cov}[w, z] $.

To offer a rough analogy for the overall tripartite approach, it is as though someone, who is accustomed to seeing the total economic activity of the United States measured by \textit{per capita} GDP (Gross domestic product divided by number of US residents), coming across an instance where economic activity in another country such as Canada is represented as total GDP, and then deciding to ``correct'' this measure, first by dividing the Canadian GDP by the number of US residents, and then by subtracting this new scaled quantity from the original total Canadian GDP in order to create a pointless new quantity.

\subsection*{Spatiotemporal extensions of the tripartite method}
The same mathematically incoherent arguments used to develop the original tripartite method have been recently applied in the development of an approach to partition biodiversity effects in space and time \citep{isbell.cowles.ea_18}. As was done in the tripartite method, \cite{isbell.cowles.ea_18} attempted to recreate what they imagined the scaled selection effect in the Price equation should look like by visual analogy. This involved creating and inserting into the change in total yield, $\Delta \phi_{_\mathbf{T}}$, without any conceivable biological or mathematical rationale, a makeshift arithmetic expression  to represent ecosystem properties as a fraction of the total  across all species, $\phi'_i / \sum_j \phi'_j$, simply as a way to visually mimic  the expression for frequency, $f'_i$, seen in the scaled Price selection effect \citep[e.g., Equation E5 Box 2 in][]{isbell.cowles.ea_18}.

Unfortunately, the jerry-rigged expressions based on this fraction fail even more spectacularly than the original tripartite method as they involve even less plausible arithmetic operations, this time between the fraction of the total property at a given site and time, $\phi'_i / \sum_j \phi'_j$, and the relative yields (proportion of monocultures), $p_i = \phi_i/M_i$ --- two distinct types of ratios which are curiously treated as indistinguishable. One of the key expressions created by arbitrarily inserting the fraction $\phi'_i / \sum_j \phi'_j$ into the the total yield was the following,
\begin{flalign} \label{eq:isbell1}
  \sum_{i=1}^n \left( \frac{\phi'_i }{\sum_j^n \phi'_j} - {p_i} \right) M_i. & &{}
\end{flalign}
The covariance term obtained by partitioning this expression is considered by \citet{isbell.cowles.ea_18} to be the equivalent of Price's selection effect, which they label the ``total insurance effect'':
\begin{flalign} \label{eq:isbell2}
  \textrm{Total Insurance Effect} &\overset{\underset{\mathrm{def}}{}}{=} n\mathrm{Cov} \left[ \left( \frac{\phi' }{\sum_j^n \phi'_j} - {p} \right), M \right]. &{}
\end{flalign}
The enigmatic variable in the above expression produced by the subtraction of two non-comparable fractions, $\left( {\phi' }/{\sum_j^n \phi'_j} - p_i \right)$ purportedly measures the ``change in dominance''. Although in \cite{isbell.cowles.ea_18} the total yield is summed over all species across different locations and points in time, we will, for simplicity's sake, assume both a single location and point in time; additional issues that arise due to their spatial and temporal approach  will not be addressed here.

The expressions created and partitioned by \citet{isbell.cowles.ea_18} using these fractions are not only completely senseless, but their use of the fraction $\phi'_i / \sum_j \phi'_j$ to represent the frequency weight $f'_i$, as seen in the scaled Price equation, only makes sense if the total yield change being partitioned is defined as $\sum_i \phi'_i M_i$. To understand why, we can backtrack from the total insurance effect, Eq. \eqref{eq:isbell2}, to determine what type of quantity is being partitioned.

If \eqref{eq:isbell2} is equivalent to the scaled Price selection effect (putting aside for now its inappropriate  arithmetic), then the fraction $\phi'_i / \sum_j \phi'_j$ would need to be equivalent to the observed frequency weight $f'_i = w_i/\sum_j w_j$, which will be true only in the (unlikely) case where the initial property values of all species are identical. Also, since $p_i=1/n$ in most BEF experiments, we can overlook the fact that the initial proportion $p_i$ is being incorrectly treated as though it were equivalent to the average weight $1/n$. This gives us
\begin{flalign} \label{eq:isbell3}
n\mathrm{Cov} \left[ \left( \frac{\phi' }{\sum_j^n \phi'_j} - \frac{1}{n} \right), M \right] &= \frac{1}{\overline{\phi'}} \; \mathrm{Cov} \Big[ \left( \phi' - \overline{\phi'} \, \right), M \Big]. &{}
\end{flalign}
After some rearrangement the RHS of \eqref{eq:isbell3} can be shown to be
\begin{flalign} \label{eq:isbell4}
 \frac{1}{\overline{\phi'}} \; \mathrm{Cov} \Big[ \left( \phi' - \overline{\phi'} \, \right), M \Big] &= \frac{1}{\overline{\phi'}} \; \mathbb{E}\Big[ \left( \phi' - \overline{\phi'} \, \right) M \Big].  &{}
\end{flalign}
By reversing the steps to derive the scaled selection effect seen in Eqs. \eqref{eq:selection_total}-\eqref{eq:selection_ave}, we can obtain the unscaled version of \citeauthor{isbell.cowles.ea_18}'s selection effect from Eq. \eqref{eq:isbell4} above:
\begin{flalign} \label{eq:isbell5}
  n\mathrm{Cov} \Big[ \left( \phi' - \overline{\phi'} \, \right), M \Big] &=  n\mathbb{E}[ \phi' \, M ] - n\mathbb{E}[ \overline{\phi'}\, M  ] &{} \notag \\
  &= n\mathbb{E}[ \phi' \, M ] - n\,\overline{\phi'}\,\overline{ M }  &{} \notag \\
  &= n\mathrm{Cov} [  \phi' , M ].  &{}
\end{flalign}

Since $n\mathrm{Cov} [  \phi',  M ] = n\mathbb{E}[  \phi'  M ] - n\,\overline{\phi'}\,\overline{ M }$, the  total system change that corresponds to \citeauthor{isbell.cowles.ea_18}'s' version of the Price selection effect will be given by
\begin{flalign} \label{eq:isbell6}
 n\mathbb{E}\Big[ \phi'\, M  \Big] &= \sum_{i=1}^n  \phi_i' \, M_i. &{}
\end{flalign}
Here the observed property of species $i$, $\phi'_i$, can play the proxy role of fitness, but only if all the initial property values are not only equal, but also all equal to 1. From \eqref{eq:isbell6} we see how the assumptions underlying  \citeauthor{isbell.cowles.ea_18}'s  total insurance effect imply that the total property or yield observed is $\sum_i^n  \phi_i' \, M_i$,  thus rendering the total yield an inscrutable quantity measured in squared units (e.g., $\mathrm{biomass}^2$). Thus, the end result of the whole spatiotemporal partitioning endeavor appears to have only been to multiply by several fold the number of meaningless new partitions that measure nonexistent phenomena \citep[Box 3 in][]{isbell.cowles.ea_18}.

\section*{Interpreting the BEF Price equation: a geometric approach} \label{BEF_price}
\subsection*{The BEF Price equation}
If we stick to our assumption that ecosystem properties in species monocultures are linear functions of species abundance then it is easy to demonstrate that the original LH partitioning would be better understood as a partitioning of the transformational component of the Price equation, and furthermore, that the LH selection term, if not strictly identical to the variational selection term in the Price equation, at least represents a similar and comparable type of  effect. In other words, while the Price equation selection term represents the proportion of \textit{variational} change that is due to selection, Loreau and Hector's selection term represents that proportion of the \textit{transformational} change that contributes or augments the variational selection effect. Below we offer a more reasonable way of interpreting BEF experiments using the Price equation.

Recall that the total variational change in a system is simply $\phi_{_\mathbf{V}} = \sum_{i=1}^n w_i \phi_i $. In a system where each species is constrained by a finite niche volume, the final (expected) state of a species after a single time step can help determine its expected discrete change or growth factor. The initial state (size) of the community will determine the fraction of niche volume still available to be filled, and thus helps determine the  growth or fitness equivalent, $w_i$, of each species. If the abundance of species $i$ in a monoculture after a time step is $K_i$, then the fitness of each species can be represented by $w_i=K_i/\sum^n_{j=1}x_j$ (or alternatively, $w_i=M_i/(a_i\sum^n_{j=1}x_j)$), where $\sum^n_{j=1}x_j$  is the initial community size or niche occupancy. That is, the final expected state in a time step divided by the initial state of the occupied niche gives the multiplicative growth factor by which a given species $i$ would be expected to grow had the initial community (or occupied niche volume) been solely composed of a monoculture of species $i$.

One way of interpreting the variational term of the Price equation for BEF studies is as follows:
\begin{flalign} \label{eq:geometric_var}
  \phi_{_\mathbf{V}} = \sum_{i=1}^n w_i \phi_i &= \sum_{i=1}^n \frac{K_i}{\sum_{j=1}^n x_j} \, (a_i x_i) &{} \notag \\
  &= \sum_{i=1}^n K_i \, a_i \, \frac{x_i}{\sum_{j=1}^n x_j} &{} \notag \\
  &= \sum_{i=1}^n M_i \rho_i, &{}
\end{flalign}
where the initial state is given by $\rho_i$, the abundance fraction of species $i$ in the initial community, or $i$'s realized proportion of the total occupied habitat. Note that since $M_i \rho_i$ is equal to the $i$th species property after variational growth, we have $\phi_{\mathbf{V}i} = M_i \rho_i = M_i p_i$, and so the fraction of species $i$ in the initial community will also be the proportion of species $i$'s total monoculture yield attained after a single time step, $p_i = \rho_i$. This allows us to write the total variational change as $ \phi_{_\mathbf{V}} = \sum_{i=1}^n M_i \, p_i$. Since the state of the ecosystem property in Eq. \eqref{eq:geometric_var} is now represented not as the raw value, but as the realized fraction of the monoculture yields, the monoculture yields themselves can serve as the modifying or growth parameter that indicates the expected ecosystem state or value of each species in a time step.

Similarly, if the observed aggregate ecosystem property after a time step is $\phi_{_\mathbf{obs}} = \sum_{i=1}^n \phi'_i $, then the transformational component of the system change, $\Delta \phi_{_\mathbf{T}} = \phi_{_\mathbf{obs}}-\phi_{_\mathbf{V}} $ can be represented as
\begin{flalign} \label{eq:geometric_T}
  \Delta \phi_{_\mathbf{T}}  = \phi_{_\mathbf{obs}}-\phi_{_\mathbf{V}} &= \sum_{i=1}^n \phi'_i -M_i p_i &{} \notag \\
  &= \sum_{i=1}^n M_i (\frac{\phi'_i}{M_i} - p_i) &{} \notag \\
  &= \sum_{i=1}^n M_i (p'_i - p_i). &{}
\end{flalign}
Since $p'_i$ is the final observed proportion of species $i$'s monoculture yield, this gives us $ \Delta \phi_{_\mathbf{T}} = \sum_{i=1}^n M_i \Delta p_i$.

The BEF equivalent Eq. \eqref{eq:Price1} showing the total variational and transformational component of ecosystem change is thus
\begin{flalign} \label{eq:geometric_obs}
  \phi_{_\mathbf{obs}} = \phi_{_\mathbf{V}}  + \Delta \phi_{_\mathbf{T}}  &= \sum_{i=1}^n M_i p_i + \sum_{i=1}^n M_i \Delta p_i. &{}\notag \\
  &= n\mathbb{E}[Mp]  + n\mathbb{E}[M \Delta p]. &{}
\end{flalign}
In BEF research the variational component, $\phi_{_\mathbf{V}}$, lets us predict the aggregate ecosystem property expected in mixtures based on how each species grows individually in monocultures. However it is the transformational change, $\Delta \phi_{_\mathbf{T}}$ that is critical in BEF studies since it quantifies the degree to which the observed aggregate property in mixtures departs from that expected based on individual species growth, and as a result, gives us a measure of how biodiversity in mixtures may be affecting ecosystem functioning. For this reason the quantity $\Delta \phi_{_\mathbf{T}}=n\mathbb{E}[M \Delta p]$ measured in experiments is known as the net biodiversity effect \citep{loreau.hector_01}.  It is the net biodiversity effect that is further partitioned by the Loreau-Hector method into selection and complementarity effects in order to infer the relative contributions of species selection/competition and niche complementarity or partitioning to the overall net biodiversity effect \citep{loreau.hector_01,pillai.gouhier_18}.

The total variational component, $\phi_{_\mathbf{V}} = n\mathbb{E}[Mp] $, can be considered the sum of changes that are attributable to both selection, and to the average growth of the system as a whole, $ n\mathbb{E}[Mp] = n\mathrm{Cov}[M, p] + n\overline{M}\overline{p} $.  Since the classical Price equation is only concerned with that portion of the variational growth strictly attributable to selection, we can subtract the average expected growth of the system from both sides of Eq. \eqref{eq:geometric_obs}, and then, by recalling that $\phi_{_\mathbf{obs}}=\sum_i M_i \, p'_i$, we can rewrite Eq. \eqref{eq:geometric_obs} as the Price equation quantifying the  \textit{total} change in the system due to the selection effects and transformational changes:
\begin{flalign} \label{eq:BEF_price_total}
  \sum_{i=1}^n \left( M_i\,p'_i  - \overline{M}\, p_i \right)  &= n\mathrm{Cov}\left[M, p\right] + n\mathbb{E}\left[M \Delta p\right]. &{}
\end{flalign}
Dividing by number of the species, $n$, gives the the \textit{total per species} version of the BEF Price equation
\begin{flalign} \label{eq:BEF_price_percap}
  \overline{M} \sum_{i=1}^n \left( \frac{M_i}{\sum_j M_j} \,p'_i  - \frac{1}{n}\, p_i \right)  &=  \mathrm{Cov}\left[M, p\right] + \mathbb{E}\left[M \Delta p\right]. &{}
\end{flalign}
Equation \eqref{eq:BEF_price_percap} is the BEF equivalent of the Price equation in evolutionary biology measuring total per group change in the trait value, with proportion of monoculture yield replacing trait $z$  as the property of interest being measured: $\overline{w}\Delta\overline{z} = \mathrm{Cov}[w, z] + \mathbb{E}[w \Delta z]$.

Given that the monoculture yields operate as proxies of fitness, we can rearrange Eq. \eqref{eq:BEF_price_percap} into the form of the Price equation describing the \textit{average per species} change change in property $p$,
\begin{flalign} \label{eq:BEF_price_ave}
  \sum_{i=1}^n \left( \frac{M_i}{\sum_j M_j} \, p'_i - \frac{1}{n} \, p_i\right)  &= \frac{1}{\overline{M}} \, \mathrm{Cov}[M, p] + \frac{1}{\overline{M}} \, \mathbb{E}[M \Delta p]. &{}
\end{flalign}
The LHS of Eq. \eqref{eq:BEF_price_ave} gives the change in the average property of the system (proportion of realized monoculture yields, $p_i$), and as such, can be represented as $\Delta \overline{p}  =\left(\, \overline{p'}-\overline{p} \, \right) $. This is the BEF equivalent of the evolutionary version of the Price equation, $\Delta \overline{z}  =\left(\, \overline{z'}-\overline{z} \, \right) =  \frac{1}{\overline{w}}\mathrm{Cov}[w, z] + \frac{1}{\overline{w}}\mathbb{E}[w \Delta z]$.

In the special case where the transformational change vanishes ($ \mathbb{E}[M \Delta p] =0$), both Eqs. \eqref{eq:BEF_price_total} and \eqref{eq:BEF_price_ave} will give expressions for the the total variational selection effect $n\mathrm{Cov}[M, p]$ and the average variational selection $\frac{1}{\overline{M}} \, \mathrm{Cov}[M, p]$, respectively. Both expressions, unlike the dominance effect represent true BEF versions of the classic selection effect. The true average or scaled BEF version of the scaled variational selection term from evolutionary biology is
\begin{flalign} \label{eq:BEF_selection_ave}
  \sum_{i=1}^n \left( \frac{M_i}{\sum_j M_j} - \frac{1}{n} \, \right)p_i  &= \sum_{i=1}^n  \left( f_i' -  \frac{1}{n}  \right)p_i  =  \frac{1}{\overline{M}} \, \mathrm{Cov}[M, p]. &{}
\end{flalign}
where $f'_i$ is the final frequency weight of species $i$'s property value in mixture.

If, on the other hand, all the initial proportions, $p_i$, are equal (as is the case in BEF experiments) then the selection term vanishes and the Price equation reduces to effects arising from transformational changes alone (e.g., species interactions). Partitioning the total transformational change, $ n\mathbb{E}\left[M \Delta p\right]$ in Eq. \eqref{eq:BEF_price_total} gives us the the total transformational selection effect given by the LH selection term $n \mathrm{Cov}\left[M \Delta p\right]$, whereas partitioning the scaled transformational term in Eq. \eqref{eq:BEF_price_ave} gives the average or scaled  version of the LH selection term,
\begin{flalign} \label{eq:BEF_T_selection_ave}
  \sum_{i=1}^n \left( f'_i  - \frac{1}{n}  \right)  \Delta p_i &= \frac{1}{\overline{M}} \, \mathrm{Cov}[ M, \Delta p]. &{}
\end{flalign}
Although the scaled version of the LH selection effect may not be particularly useful in BEF research, we can still note that it bears no meaningful resemblance to the ``dominance effect'' (Eq. \eqref{eq:dominance_LH}) obtained from the tripartite partitioning.

\subsection*{Vector interpretation of the BEF Price equation}
Representing the effects described by the BEF Price equation, as well as the Loreau-Hector partitioning of the transformational term, as vectors provides a visually intuitive geometric interpretation of the ecosystem changes being scored \citep{pillai.gouhier_18}. We can picture the state of an $n$-species ecosystem at any given time as the coordinates in a state space, where each axis represents a species ecosystem property contribution. For example the state of the system expected due to variational growth can be represented by the coordinate $\Phi_{V} = (\phi_{V1}, \phi_{V2},\dotsc,\phi_{Vn})$, and the observed state by the coordinate $\Phi_{\mathbf{obs}} = (\phi'_{1}, \phi'_{2},\dotsc,\phi'_{n})$.

\begin{figure}
  \centering
  \includegraphics[scale=0.42]{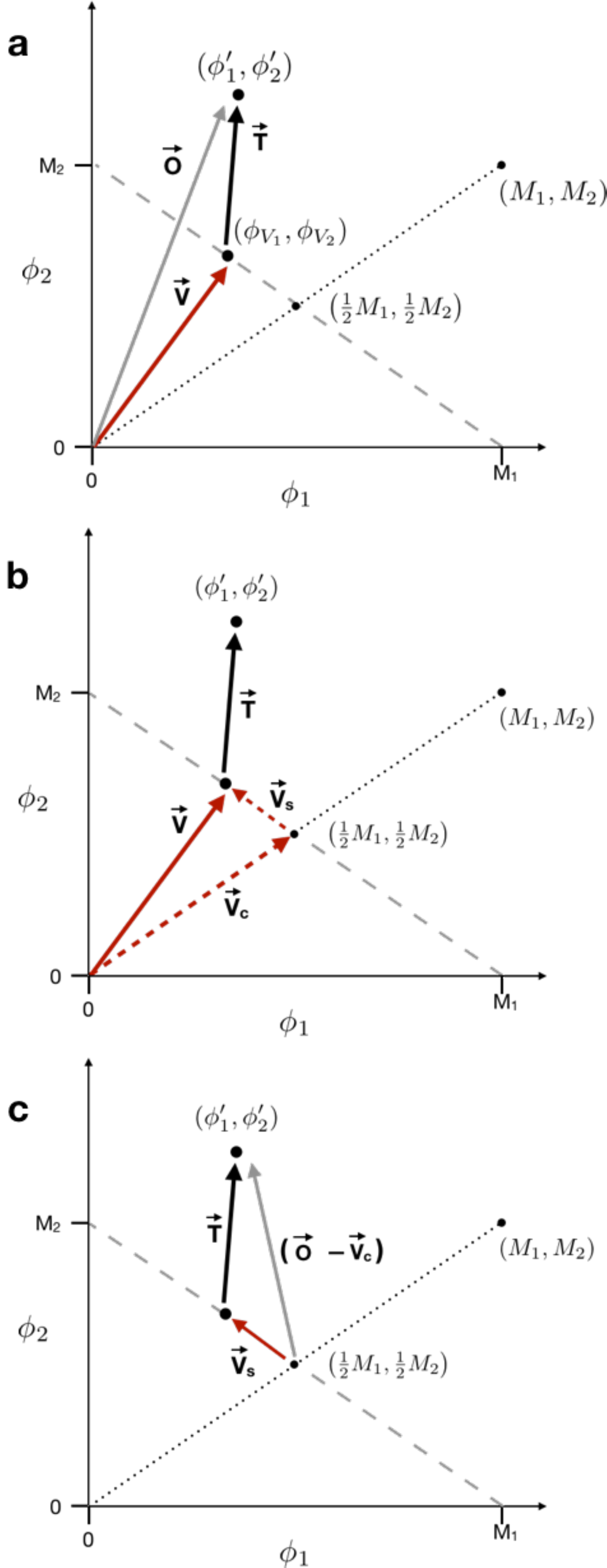}
  \caption{Vector interpretation of the BEF Price equation. The state of the ecosystem can be shown as coordinates in an ecosystem state space. Example shown for two-species system. (a) Positional vector of the observed ecosystem state, $\mathbf{O}$, shown as the sum of two vectors: $\mathbf{V}$ giving the state expected due to variational growth, and $\mathbf{T}$, showing transformational change giving the observed departure from the expected state. (b) $\mathbf{V}$ resolved into components $\mathbf{V_c}$, along the niche partitioning axis, and $\mathbf{V_s}$, falling along the monoculture simplex from the centroid to the expected state. (c) Total ecosystem change in BEF Price equation is the sum of shifts along $\mathbf{V_s}$ and  $\mathbf{T}$ vectors: $( \mathbf{O} - \mathbf{V_c}) = \mathbf{V_s} + \mathbf{T}$.}
  \label{Fig:fig1}
\end{figure}
Each point in the ecosystem space can also be represented as a positional vector from the origin to that point. Thus we can represent the expected state of the system due to variation in individual species growth rates (the variational component of change) as the positional vector  $\mathbf{V} = \big\langle \phi_{V1}, \phi_{V2},\dotsc,\phi_{Vn} \big\rangle$ (Fig. \ref{Fig:fig1}a). Similarly, the positional vector to the observed state, $\Phi_{\mathbf{{obs}}}$ can be represented by the vector $\mathbf{O} = \big\langle \phi'_1, \phi'_2,\dotsc,\phi'_n \big\rangle$. The transformational changes in the system between the expected state $\Phi_{V}$, and the observed state $\Phi_{\mathbf{{obs}}}$ is represented by the displacement vector $\mathbf{T}$, where $\mathbf{T} = \mathbf{O} - \mathbf{V} = \big\langle (\phi'_1 -\phi_{V1}), (\phi'_2-\phi_{V2}), \dotsc, (\phi'_n-\phi_{Vn}) \big\rangle$. Thus the observed ecosystem state changes due to transformational and variational changes in the Price equation (Fig. \ref{Fig:fig1}a), as was seen in Eq. \eqref{eq:Price1} ($\phi_{_\mathbf{obs}} = \phi_{_\mathbf{V}}  + \Delta \phi_{_\mathbf{T}}$), can be represented as
\begin{flalign} \label{eq:vector_V_T}
\mathbf{O}  &= \mathbf{V} + \mathbf{T}. &{}
\end{flalign}
The total ecosystem change along any vector is simply the sum of all species' ecosystem changes along that vector. Consequently, the total ecosystem property change along $\mathbf{V}$ and $ \mathbf{T}$ are given by  $ \phi_{_\mathbf{V}} = \sum_{i} \phi_{Vi}$ and $ \Delta \phi_{_\mathbf{T}} = \sum_{i} (\phi'_i - \phi_{Vi})$, respectively.

We can resolve the variational vector $\mathbf{V}$ into two components, $\mathbf{V} = \mathbf{V_s} +\mathbf{V_c}$: one component falling on the simplex (or hyperplane) connecting all monoculture yields, $\mathbf{V_s}$, and another component, $\mathbf{V_c}$, that falls along the complementarity axis, or positional vector connecting the origin to the monoculture yields $\langle M_1, M_2,\dotsc,M_n \rangle$. The first component falling along the simplex, $\mathbf{V_s} = \big\langle p_1M_1 - \overline{p}M_1, p_2M_2 -\overline{p}M_2, \dotsc, p_nM_n -\overline{p}M_n \big\rangle$, gives the total variational change due to the classic selection effect, while the second component, $\mathbf{V_c} =\big\langle \overline{p}M_1, \overline{p}M_2, \dotsc, \overline{p}M_n \big\rangle$, gives the expected growth in the system if all species had the same initial proportions (Fig. \ref{Fig:fig1}b).
\begin{figure}[h!]
  \centering
  \includegraphics[scale=0.37]{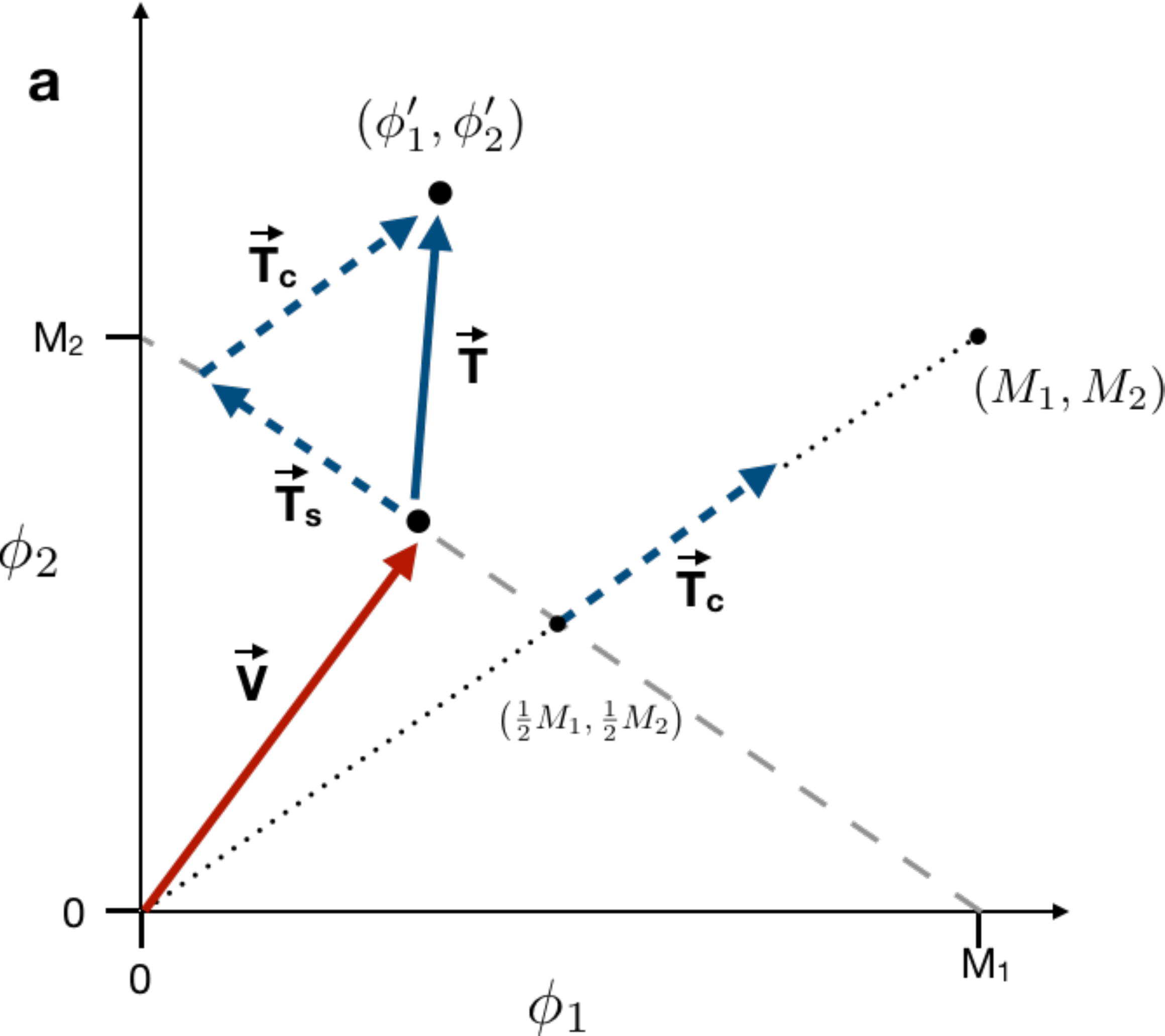}
  \caption{Resolving the transformational component $\mathbf{T}$. The vector $\mathbf{T}$ describing the net biodiversity effect can be resolved into two vectors: one  giving the LH selection effect, $\mathbf{T_s}$ along the simplex, and the other which is the LH complementarity effect, $\mathbf{T_c}$ falling along the niche partitioning axis.}
  \label{Fig:fig2}
\end{figure}
In the classic formulation of the Price equation it is only the selection component of total variational change that is considered. Thus the Price equation for total absolute change can be represented in vector form by
\begin{flalign} \label{eq:vector_price}
(\mathbf{O}- \mathbf{V_c} ) &= \mathbf{V_s} + \mathbf{T}. &{}
\end{flalign}
In both Eq. \eqref{eq:vector_price} and Fig. \ref{Fig:fig1}c, we can see how the total change described by the Price equation is given by $\mathbf{O}- \mathbf{V_c}$, which is the displacement vector from the centroid of the simplex $\frac{1}{n}(M_1, M_2, \dotsc, M_n)$ to the observed state, $ (\phi'_1, \phi'_2,\dotsc,\phi'_n)$. This the total change partitioned by the classic Price equation into selectional, $\mathbf{V_s}$,  and transformational changes, $\mathbf{T}$. Note that in standard BEF experiments where initial proportions of all species in mixture, $p$, are equal, the total ecosystem change will be due to the transformational effect $\mathbf{T}$ alone,  since $\mathbf{V_s}$ vanishes due to all variational change occurring along the complementary axis, $\mathbf{V}=\mathbf{V_c}$.

\begin{figure}[h!]
  \centering
    \includegraphics[scale=0.32]{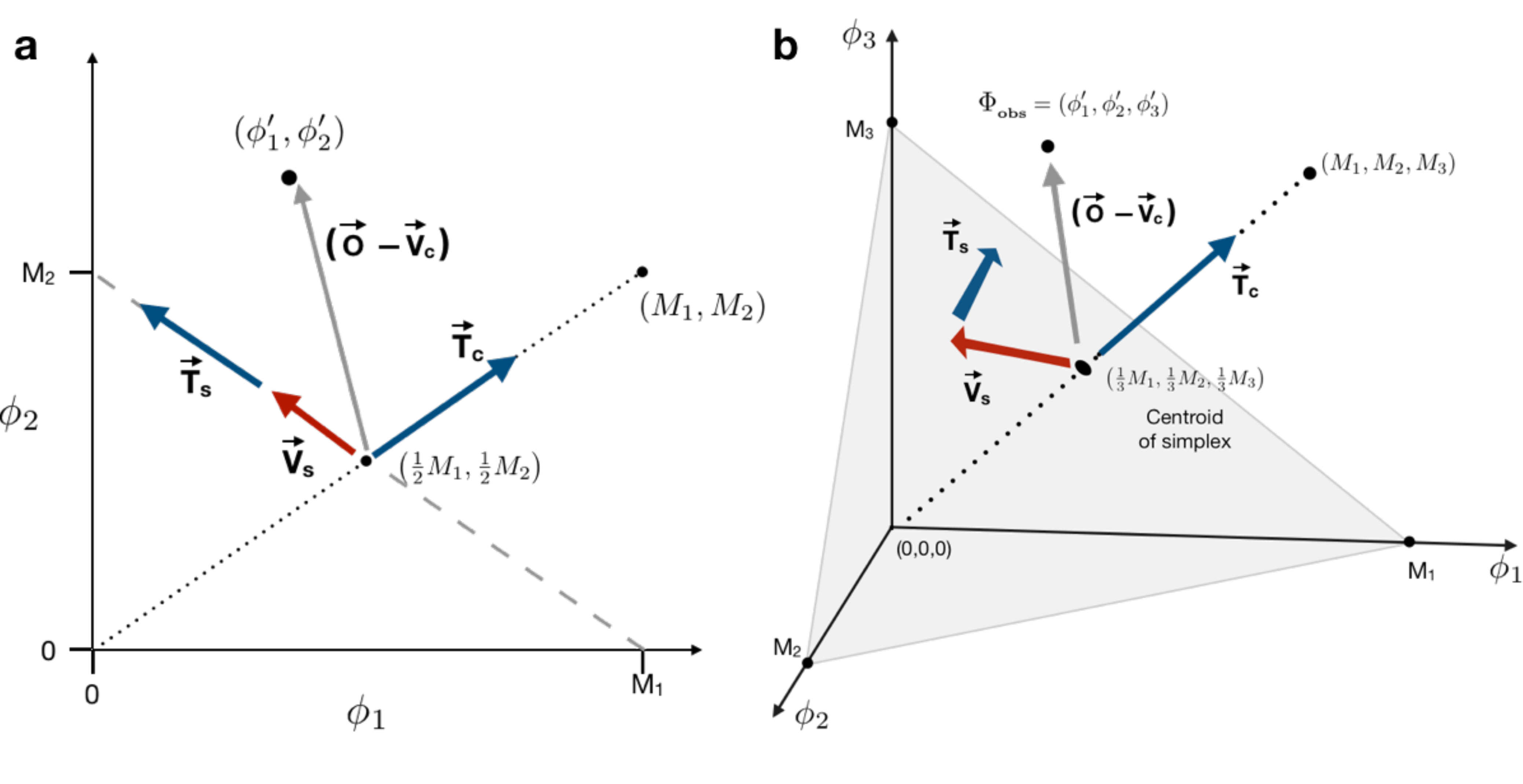}
    \caption{Component vectors for BEF Price equation. Example shown for (a) two-species system, and (b) three-species system. Note how the variational selection and LH selection vectors represent similar types of changes constrained to the same space defined by the ecosystem simplex.}
    \label{Fig:fig3}
\end{figure}

In a similar manner (Fig. \ref{Fig:fig2}), the transformational vector can be resolved into the LH-selection component vector, $\mathbf{T_s}= \big\langle (\Delta p_1 -\overline{\Delta p}\,)M_1, (\Delta p_2 -\overline{\Delta p}\,)M_2, \dotsc, (\Delta p_n-\overline{\Delta p\,)}M_n \big\rangle$, that falls along the simplex, and the LH complementarity vector, $\mathbf{T_c} =\big\langle \overline{\Delta p}M_1, \overline{\Delta p}M_2, \dotsc, \overline{\Delta p}M_n \big\rangle$, that is parallel to the complementarity axis, such that $\mathbf{T} = \mathbf{T_s} + \mathbf{T_c}$ \citep[for proofs see][Appendix S3]{pillai.gouhier_18}. Visually, it becomes very clear that, \textit{contra} Fox, the Loreau-Hector selection effect, $\mathbf{T_s}$, and the standard variational selection effect, $\mathbf{V_s}$, are comparable types of ecosystem changes constrained to the same space (Fig. \ref{Fig:fig3}). The simplex connecting the expected state of each species, had they been grown separately in monocultures, gives a space along which ecosystem or species compositional changes will represent a zero-sum game, such that any increase in a given species' contribution will come perfectly at the expense of others based on ratios given by the monoculture yields.
The expected state $\Phi_{V}$ on the simplex will be determined by initial conditions. The variational selection effect $\mathbf{V_s}$ measures shifts to this expected state from the center of the simplex, which is the ecosystem state that would be expected had the initial fraction of all species been equal. The transformational or LH selection effect $\mathbf{T_s}$ measures further departures from the expected state along the simplex, and thus is an additional selection effect that arises from such factors as species interactions in mixtures that further augment the variational selection effect (Fig. \ref{Fig:fig3}).  Accordingly, both $\mathbf{V_s}$ and $\mathbf{T_s}$ can be seen as representing contributions to a general selection effect from two different sources of ecosystem change, \textit{variational} and \textit{transformational}.

We can see in Fig \ref{Fig:fig2} how the ability to measure the LH selection effect (i.e., the transformational selection term $\mathbf{T_s}$) does not technically require the proportions of all species in the initial mixture to be equal as is standard in BEF experiments (and as is required with the tripartite method's dominance expression). The generality of the Loreau-Hector selection term is the result of it being a simple and direct extension of the classic Price equation, unlike the dominance effect, which appears to have no clear mathematical or biological link to the Price equation.

\section*{Discussion}
One of the most common justifications for using the tripartite method is the claim -- taken at face value and repeated for over a decade -- that the Loreau-Hector selection term does not represent a ``zero-sum game'', unlike the tripartite methods's dominance effect, which purportedly ``quantifies the extent to which observed species’ relative yields in mixture resemble a zero sum game'' \citep{fox_05}. As we can clearly see (Fig. \ref{Fig:fig3}), the  LH selection term represents a similar type of change as the variational selection term in the classic Price equation, and that \textit{both} types of change represent zero-sum games that are constrained to the same simplex based on fitness or growth rates (here given by total monoculture yields).

The variational selection effect measures the degree to which an initial skew in the property or trait distribution is amplified by variation in growth rates. Similarly, when all species' ecosystem-abundance relationships are linear, the Loreau and Hector transformational selection term can potentially measure how this distribution is further skewed due to the effects of competition, or other unknown factors favoring one set of species over others. However, under nonlinearity there is no real reason to further partition the transformational term as it can no longer isolate the relative contributions of different effects, and in this sense Loreau and Hector's label of ``selection effect'' for the transformational covariance term is inappropriate \citep{pillai.gouhier_18}. But under the implicit assumption of linearity, the label is completely justified, and consequently, the creation of the dominance effect was simply a nonsensical solution to a non-existent problem. In the end, the dominance effect was nothing more than a version of the Loreau-Hector selection term mangled on a procrustean bed of misapprehensions regarding the Price identity.


Although it is almost impossible to reduce the welter of incongruous lines of reasoning to a simple set of errors, we have attempted to broadly summarize below the key issues with the tripartite method:
\begin{enumerate}
\item \textbf{The claim that $\mathbb{E}[\Delta f]=0$ is characteristic of selection is \textit{prima facie} absurd.} The expression $\mathbb{E}[\Delta f]=0$ follows from the simple arithmetic property that requires fractions of some total to sum-up to one, and will thus \textit{always} apply whenever one compares \textit{any and every} set of real or imagined objects to each other, and therefore has absolutely nothing to do with ``selection, or analogous processes''. This near tautological statement is the basis and justification for the claim about zero-sum games for the tripartite method \citet{fox_05}, and its spatiotemporal extension \citet{isbell.cowles.ea_18}.

\item\textbf{The development of the argument rests on a facile \textit{visual} analogy.} The justification rests on a na\"ive claim that $\Delta p$ is comparable to $\Delta f$. Not only do $p$ and $f$ represent different types of quantities, but even the changes $\Delta$ in the two expressions are not comparable. The change in $\Delta f$ refers to the change in the frequency weight of a property \textit{relative to the average weight} of $1/n$ , while the change in $\Delta p$ refers to the change in the actual property value between its final observed and expected states. 

\item\textbf{The partitioning involved confusing a scaled and a non-scaled version of the same quantity.} The \textit{scaled} version of the variational selectional term in the classic Price equation was compared to the \textit{unscaled} version of the BEF transformational selection term (LH selection), without realizing that the functional forms of both expressions were interchangeable, or that both expressions could be put into the same functional form by a single arithmetic operation.

\item\textbf{A general arithmetic property of covariance was construed as a unique feature of the Price equation in evolutionary biology.} The ability to express a covariance term as a change in frequencies or fractions of some variable is \textit{not} a property of the Price equation or of selection in general, but a simple arithmetic property of covariance itself.

\item\textbf{There is no consideration of what plays the role of `fitness'.} Nowhere does the expected change, growth factor or fitness of individual parts or species in the system property appear to come into play. The whole point of the Price equation partitioning is to measure changes relative to expected growth or fitness.

\item\textbf{Partitioning involves calculating differences between incommensurable and incomprehensible quantities.} The very foundations of both the tripartite method and its spatiotemporal extension involve nonsensical arithmetic operations applied to unfathomable quantities that are also non-comparable, such as $\phi'_i - \frac{\phi'_i}{\sum_jp'_j}$  \citep[Eq. 10 in][]{fox_05}, or $ \phi'_i - \frac{(\phi'_i\times M_i)}{\sum_j\phi'_j}$, \citep[Eq. 5 in][]{isbell.cowles.ea_18}.

\end{enumerate}

What is particularly surprising is that despite the fact that the tripartite method has been cited hundreds of times and implemented in open-source software \citep{isbell.cowles.ea_18}, its fundamentally unsound premises have gone unnoticed for over a decade. Even worse, the same baffling lines of reasoning that were used to develop the original tripartite method have also been used in its more recent elaborations, specifically the attempt to quantify the temporal and spatial insurance effect \citep{isbell.cowles.ea_18}. As hard as it is to imagine, these recent attempts to extend the tripartite framework spatiotemporally are even more incoherent than the original.

Ultimately, the entire class of tripartite approaches all rest on the same basic misreading of the Price equation. Take, as an example, the following rationalization for the development of the spatiotemporal partitioning method:
\begin{quotation}
  \noindent Our total insurance effect term is equivalent to Price’s (1970, 1972) selection effect in evolutionary genetics. Note that Loreau and Hector's (2001) selection effect and Fox's (2005) dominance effect were both inspired by, but not equivalent to, Price's selection effect. Both these previous partitions considered the covariance between monoculture yields and species' overyielding or underyielding behaviour, rather than species' dominance, in mixtures. Species can overyield by having high yields in mixture or by having low yields in monoculture. Thus, species can overyield without dominating mixtures. In contrast, our new partition more fully isolates the covariance between monoculture yields and mixture dominance than previous partitions by including a covariance term, the total insurance effect, which quantifies the covariance between monoculture yields and a variable that depends on mixture relative abundance or biomass and that does not depend on monoculture yields. \citep{isbell.cowles.ea_18}
\end{quotation}
This remark indicates the serious and ongoing confusion as to what the selection term in the classic Price equation refers to (or even what the Price equation actually quantifies). It is certainly not a measure of some arbitrary property's ``dominance in mixtures'' relative to monoculture yields. Indeed, two species with similar initial property/trait values but different growth rates will end-up in a mixture with one of the species' traits `dominating' after one time step, but the selection effect should still be zero  -- regardless of how this so-called `dominance' covaries with monoculture yields. Nor can this `dominance' in mixtures, $\phi'_i / \sum_j \phi'_j$, even superficially serve as a measure of the observed frequency weight, $f'_i$, seen in the scaled Price equation ($f'_i = w_i / \sum_j w_j$), since the initial properties of all species cannot be assumed to be all equal (unlike initial proportions, $p_i$, in BEF experiments).

Any proper selection effect will involve measuring the shift in the distribution of an aggregate property \textit{relative} to the expected growth rates, states, or fitnesses of individual species or groups. Our geometric interpretation of ecosystem changes highlights this fact by showing how selection can be visualized as shifts constrained to the simplex connecting the expected monoculture states of all species.

These rampant and long-standing misunderstandings are quite unfortunate because the Price equation can serve as a valuable framing device for teasing-out what is known and unknown about the changes observed in a system. Indeed, awareness of the difference between variational and transformational changes in a system's evolution can allow researchers to more clearly ascertain the assumptions being made in an experiment and what possible inferences can legitimately be made regarding the observed changes. For instance, in BEF experiments the variational change in the system is the expected change in aggregate properties in a mixture under the null assumption that the growth factor in the property of all species in the mixture will simply be identical to the growth of species observed independent of each other in monocultures. The transformational component represents the changes due to whatever residual factors were not accounted for in the null assumptions based on individual species growth independent of each other.

In BEF experiments the implicit belief is that this residual category only contains factors that are associated with the effects arising from mixtures -- that is, those associated with increased biodiversity.  Since the expected growth of each species independent of all others is scored as the total yield of species obtained in monocultures, monoculture yields became the measure of expected growth, or the BEF equivalent of fitness. All of this is fine as far as it goes; however one needs to be aware of what is involved when monoculture yields are used as a proxy of fitness. Because individual species' ecosystem properties may not scale linearly with abundance in monocultures \citep[][]{yoda.kira.ea_63,westoby_84,enquist.brown.ea_98}, the residual category measured by the transformational change will now not only contain the effects due to increased diversity in mixtures, but also the effects arising from nonlinear ecosystem functioning-abundance relationships in each species' individual growth curves. Under nonlinearity Loreau and Hector's net biodiversity effect (the transformational component of change) will confound these two effects, making it effectively impossible to quantify complementarity and selection effects \citep{pillai.gouhier_18}. The fact that this critical limitation has gone unnoticed for the better part of two decades, but becomes almost self-evident in our vectorized version of the Price equation, highlights the strength of our approach.

Such a modest and focussed use of the Price identity could have allowed BEF researchers to more rigorously define the assumptions built into their experimental design, and in the process, avoid decades of misunderstanding and misdirected experimental effort.  Instead, the last decade has seen a proliferation in the pretentious and obfuscating use of the phrase ``Price equation partitioning'' for almost any application of the covariance operator in the analysis of biodiversity data, particularly for methods employing spreadsheet-style bookkeeping calculations when making descriptive comparisons between communities \citep{fox_06, fox.harpole_08, fox.kerr_12, bannar-martin.kremer.ea_18}

Overall, the tripartite method and its extensions represent a unique and unprecedented style of argumentation, involving the use of senseless arithmetic operations, justifications motivated by false analogies,  deductions based on superficial resemblances of incommensurable variables, as well as logical and mathematical non-sequiturs.  However, despite these critical flaws, the method has been used to analyze biodiversity data for nearly a decade-and-a-half, most recently as a means to partition biodiversity data spatiotemporally \citep{garcia.bestion.ea_18}. Nevertheless, despite its previous misuses, the Price equation may still provide a useful tool in the design and interpretation of ecosystem experiments. Our mathematical derivation and visual depiction of a vectorized version of the Price equation in terms of variational and transformational change can help to clarify the important caveats and limitations associated with the statistical decomposition and analysis of ecosystem change, in addition to helping us interpret the selection and complementary effects.

\section*{Acknowledgements}
This  research  was  supported  by  a  grant  from  the  National  Science  Foundation  (CCF-1442728) to TCG

\bibliography{pillai_gouhier_2018_BEFtripart_biblio}

\bibliographystyle{ecology}

\end{document}